\documentclass[aps,prl, twocolumn, showpacs]{revtex4}
\usepackage{amsfonts}
\usepackage{graphicx}
\usepackage{dcolumn}
\usepackage{bm}

\begin{document}
\preprint{preprint}
\title{Exciting a $d$-density wave in an optical lattice with driven tunneling}
\author{A. Hemmerich}
\affiliation{Institut f\"{u}r Laser-Physik, Universit\"{a}t Hamburg, 
Luruper Chaussee 149, 22761 Hamburg, Germany}
\author{C. Morais Smith}
\affiliation{Institute for Theoretical Physics, Utrecht University, 3508 TD Utrecht, The Netherlands}
\date{\today}

\begin{abstract}
Quantum phases with unusual symmetries may play a key role for the understanding of solid state systems at low temperatures. We propose a realistic scenario, well in reach of present experimental techniques, which should permit to produce a stationary quantum state with $d_{x^2-y^2}$-symmetry in a two-dimensional bosonic optical square lattice. This state, characterized by alternating rotational flux in each plaquette, arises from driven tunneling implemented by a stimulated Raman scattering process. We discuss bosons in a square lattice, however, more complex systems involving other lattice geometries appear possible. 
\end{abstract}


\maketitle
Solid state systems at low temperatures may exhibit exotic quantum phases, which involve order parameters with unusual symmetries. Various researchers have considered order parameters with $d$-wave symmetry \cite{Aff:88, Zha:90,Hsu:91,Nay:00, Cha:01} with the hope of shedding light on the yet unexplained pseudogap phase in cuprate superconductors. Quantum states with $d_{x^2-y^2}$-wave symmetry (staggered flux states), characterized by alternating rotational flux in each plaquette, have been shown to arise in the context of the two-dimensional Hubbard-Heisenberg model \cite{Aff:88}, or in the context of the $t-J$ model \cite{Zha:90}. Unfortunately, decisive experiments, which could clarify the role of such unconventional symmetry broken phases, are hampered by the effects of inevitable disorder in solid state systems. 

In recent years, atomic or molecular quantum gases \cite{QuantumGases} loaded into the perfectly periodic light-shift potentials of optical lattices have emerged as synthetic laboratory systems which permit to experimentally simulate basic models of low temperature physics while keeping all system parameters under control \cite{Lew:06}. A seminal example is the observation of the superfluid-insulator phase transition within the Bose-Hubbard-model \cite{Fis:89, Jak:98, Gre:02}. Ever since there have been rapidly growing activities to engineer optical lattice models of more involved cases like paired fermionic systems \cite{Molecules, Koe:06}, quantum Hall systems \cite{QuantumHall}, or to study unusual lattice geometries like Kagom\'e lattices \cite{San:04}. According to a recent proposal \cite{Buc:05}, quantum states with unconventional order could be implemented in an optical lattice of bosons if their internal structure allows for additional engineered interactions. Unfortunately, the existing proposals to extend the scope of optical lattices to the exciting realms beyond the conventional Hubbard model are experimentally extremely demanding or even involve requirements which cannot be met with state of the art technology.

In this article we discuss a simple realistic scenario which allows for the experimental implementation of an order parameter with $d$-wave symmetry in a two dimensional square optical lattice of simple polarizable bosons. Our proposal relies on the existence of a metastable stationary solution of the Gross-Pitaevski (GP) equation with $d_{x^2-y^2}$-wave symmetry. In order to excite this state, we propose a method based upon Raman scattering. Apart from a non-zero polarizability, no internal structure of the bosons is required, a circumstance which should dramatically facilitate experimental investigations. We exploit the interference between two appropriate light fields with slightly different frequencies, which produces a two-dimensional square array of microscopic rotors. These rotors generate angular momentum with alternating rotational direction for adjacent plaquettes. In the framework of the Bose-Hubbard-model, we show that this Raman coupling leads to driven ring-like tunneling characterized by anisotropic time-dependent hopping. This opens up a new dimension of phase space, possibly giving rise to new quantum phases, accessible in the clean environment of an optical-lattice system. Here, we restrict ourselves to bosonic atoms in a square lattice geometry, however, more involved scenarios are in reach of present experiments. For example, a molecular optical lattice could be used to access the BCS regime \cite{Koe:06}. $D$-density waves subjected to triangular or quasi-periodic lattices should provide exciting new options to study frustration phenomena \cite{Lew:06}. 

Let us consider solutions of the time-independent GP-equation in an external potential $V_{\rm trap}( \mathbf r)$ 
\begin{equation}
\label{GP} 
\Big[\frac{- \hbar^{2}}{2m}\Delta  +  V_{\rm trap}( \mathbf{r})  + g \rho( 
\mathbf{r})\Big] \psi( \mathbf{r}) =\mu \, \psi( \mathbf{r}) 
\end{equation}
of the form  $\psi( \mathbf{r}) = \sigma( \mathbf{r})\phi( \mathbf{r})$, with $\rho( \mathbf{r}) \equiv |\psi( \mathbf{r})|^2$ the particle density and $\mu$ the chemical potential. The collision parameter $g \equiv 4 \pi \hbar^{2} a /m$, with $m$ the atomic mass and $a$ the $s$-wave scattering length, is positive for repulsive collisions. Upon use of the eigenfunction of the kinetic energy $\phi(x,y) \equiv e^{i \theta/2}\sin(kx) + e^{-i \theta/2}\sin(ky)$, i.e., $- (\hbar^{2}/2m) \Delta \phi = E_{\rm rec} \phi$ with the recoil energy $E_{\rm rec} \equiv \hbar^{2} k^{2}/2m$, Eq.\ (1) becomes $- (\hbar^{2}/2m) [ \Delta \sigma / \sigma +  2 \mathbf{ \nabla} \ln(\sigma)  \mathbf{\nabla} \ln(\phi) ] + E_{\rm rec} +  V_{\rm trap}  + g \rho =  \mu$. Assuming a positive, slowly varying envelope function $\sigma( \mathbf r)$ such that $|  \mathbf{\nabla} \sigma| \ll k \sigma $ and $|\Delta \sigma| \ll k^{2} \sigma$, one may approximate $| \Delta \sigma/\sigma  +  2   \mathbf{\nabla} \ln(\sigma) \mathbf{\nabla} \ln(\phi)| \leq  |\Delta \sigma / \sigma| +  2 |\mathbf{\nabla} \ln(\sigma)||   \mathbf{\nabla} \ln(\phi)| \leq |\Delta \sigma / \sigma | +  2k |  \mathbf{\nabla} \ln(\sigma)| \ll 3k^{2}$ and thus $| \Delta \sigma /\sigma +  2   \mathbf{\nabla} \ln(\sigma) \mathbf{\nabla} \ln(\phi)| \ll E_{\rm rec}$. Consequently, the expression between square brackets may be neglected as compared to $E_{\rm rec}$. This leads to $E_{\rm rec} +  V_{\rm trap}  + g \rho =  \mu$ with the possible solution 
\begin{equation}
\label{solution} 
\mu = E_{\rm rec}\,\,,\,\,\, V_{\rm trap} = - g \rho \,\, , 
\end{equation}
where $\rho = \sigma^2|\phi|^2$ and $|\phi|^2 =  \sin^{2}(kx)+\sin^{2}(ky) 
+ 2\cos(\theta)\sin(kx)\sin(ky)$.
With $\phi = |\phi| e^{iS(x,y)}$  the phase $S(x,y)$ becomes
\begin{eqnarray}
\label{eikonal} 
S(x,y) = \arctan\left[\tan\left(\frac{\theta}{2}\right)\,
\frac{\sin(kx)-\sin(ky)}{\sin(kx)+\sin(ky)}\right]
\nonumber \\ 
+ \, \, \frac{\pi}{2}  \, \, {\rm sgn }[\Im(\phi)] \, (1 - {\rm sgn} [\Re(\phi)]),
\end{eqnarray}
where sgn$[z]$ denotes the sign of $z$. The particle current density $\mathbf{j}(x,y) \equiv |\phi(x,y)|^{2} \, \mathbf{v}(x,y)$ connected to the velocity field $\mathbf{v}(x,y) \equiv (\hbar/m)  \mathbf{\nabla} S(x,y)$ evaluates to $\mathbf{j}(x,y) = (\hbar/m) \, \sin(\theta) \, \mathbf{\nabla} \times \mathbf{\hat z}  \, \sin(kx) \sin(ky)$, i.e., it is purely vortical (see Fig.\ 1(a)). In the case $\sin(\theta) = 1$, where $\mathbf{j}$ is maximal, the local particle density scales as $ |\phi(x,y)|^{2} = \sin^2(ky) + \sin^2(kx)$ (see Fig.\ 1(b)) and the velocity field satisfies
\begin{eqnarray}
\label{vorticity} 
\mathbf{v}(x,y) &=& \, \frac{\hbar}{m} \, \, \frac{ \mathbf{\nabla} \times \mathbf{\hat z}  \, \sin(kx) \sin(ky)}{\sin^2(ky) + \sin^2(kx)}  \,  \, ,
\nonumber \\   \\  \nonumber
\mathbf{\nabla} \times \mathbf{v}(x,y) &=& \sum_{n,m \in \mathbb{Z}} (-1)^{n+m} \delta \left(x-n\frac{\pi}{k},y-m\frac{\pi}{k}\right).
\end{eqnarray}

According to Eq.\ (\ref{vorticity}), the structure of $\mathbf{v}(x,y)$  is that of a vortex-anti-vortex lattice with vortex-filaments at positions $kx,ky \in \pi \, \mathbb{Z}$. At the potential minima (white regions in Fig.\ 1(b)) the particle density is maximal and $\mathbf{v}(x,y)$ exhibits quadrupole symmetry. The vortices are pinned at the potential maxima of the lattice (black regions), in accordance with results obtained for plain vortex lattices prepared in large scale traps and subsequently exposed to an optical lattice potential \cite{Rei:05, Tun:06}. The kinetic energy and angular momentum per particle are $E_{\rm kin} = (2/\pi)E_{\rm rec}$ and $L =8 \hbar/\pi^2$, respectively. The particle density $\rho(\mathbf{r})$ tends to zero within the vortex cores, which have a radius on the order of $R_{\rm core} \approx 1/k$. Defining the healing length as $\xi \equiv (8\pi a \bar \rho)^{-1/2}$ with mean density $\bar \rho$, upon use of Eq.\ (\ref{solution}) one may write $(k \xi)^2 = E_{\rm rec}/g \bar \rho= E_{\rm rec}/ \bar V$ with mean potential well depth $\bar V$. Thus, the condition that the core size exceeds the healing length $\xi < R_{\rm core}$ is equivalent to $E_{\rm rec} <  \bar V$.

\begin{figure}
\includegraphics[scale=0.22]{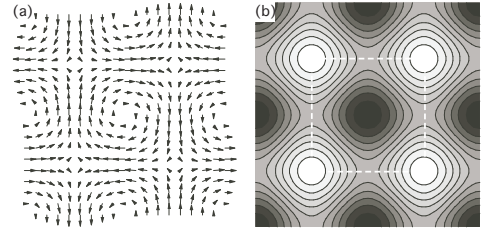}
\caption{\label{fig.1} (a) Flux density and (b) local particle density for $\sin(\theta) = 1$. Black (white) indicates low (high) particle density. The dashed white square defines the position and size of the plaquettes used to build up the lattice.} 
\end{figure}

The stability of solutions $\psi$ of Eq.\ (\ref{GP}) may be considered via the grand canonical potential $K[\psi_{\varepsilon}] \equiv \int d^{3}r [\psi_{\varepsilon}^{*}\hat T \psi_{\varepsilon} + (V_{\rm trap}-\mu) |\psi_{\varepsilon}|^2 + g |\psi_{\varepsilon}|^4/2 ]$ for $\psi_{\varepsilon} \equiv \psi + \varepsilon \chi$ with the kinetic energy operator $\hat T$, $\varepsilon \in \mathbb{R}$, and an arbitrary normalized wavefunction $\chi$. Use of Eq.\ (\ref{solution}) yields $\frac{\partial}{\partial \varepsilon} K[\psi_{\varepsilon}]_{\varepsilon=0} = 0$ and $\frac{\partial^2}{\partial \varepsilon^2} K[\psi_{\varepsilon}]_{\varepsilon=0} = \int d^{3}r  [ 2 \chi^{*} (\hat T-E_{\rm rec}) \chi - V_{\rm trap}\,(e^{-iS} \chi + e^{iS} \chi^{*})^2 ]$. This integral is evaluated on each plaquette by expanding $\chi$ in a Fourier series up to second order (higher order terms do not contribute because $V_{\rm trap}(x,y)$ is of second order). Hence, by solving an eigenvalue problem with finite dimensions we find that $\psi$ is metastable in the sense that it establishes a true local energy minimum if $\bar V > 3 E_{\rm rec}$.

The experimental realization of $V_{\rm trap}$ as required by Eq.\ (\ref{solution}) may be achieved by implementing a light shift potential obtained from the superposition of two optical standing waves arranged parallel to the $x$- and $y$-axes, with linear polarization parallel to the $z$-axis \cite{Hem:91}. If $g > 0$, the trap potential satisfies $V_{\rm trap}\leq 0$  and may thus be confining, which corresponds to adjusting a negative detuning of the optical standing waves with respect to the relevant atomic resonance. Available laser technology allows for lattice constants $\lambda /2 = \pi / k$  in a wide range between a few hundred nm and several $\mu$m. The $d_{x^2-y^2}$-wave solution $\psi$ does not represent the ground state and thus, in addition to the lattice potential, an experimental procedure is required to excite it. In the following, a method based upon stimulated Raman scattering is discussed for the case of polarizable bosons in a two-dimensional square optical lattice. We extend an approach developed for vortex excitation \cite{VortexExcitation}, such that we may apply angular momentum quanta $\hbar$ with alternating signs to the $\lambda/2$-sized square plaquettes. Let us consider a linearly polarized bichromatic light-field $\mathbf E( \mathbf r, t) = \hat z \, [E_1( \mathbf r, t)+E_2( \mathbf r, t)]$ defined by $E_1( \mathbf r, t)  \equiv  A_1\, e^{i \omega t} |\phi(x,y)|e^{i S(x,y)}$ and $E_2( \mathbf r, t)  \equiv  A_2\, e^{i (\omega+\Omega) t} |\phi(x,y)|e^{-i S(x,y)}$ with real positive amplitude coefficients $A_1$, $A_2$, frequencies $\omega$, \mbox{$\omega+\Omega$}, and $|\phi(x,y)|$ and $S(x,y)$ as in Eq.\ (\ref{eikonal}). Here, $|\Omega| \ll \omega$ is assumed such that both fields, despite their different frequencies, in good approximation share the same eikonal $S(x,y)$. The total intensity of the light field $I(x,y,t) \equiv  \mathbf E( \mathbf r, t)  \mathbf E^{*}( \mathbf r, t) = I_L(x,y) + I_R(x,y,t)$ (averaged over the optical period $2 \pi/\omega$) consists of a stationary term $I_L(x,y)$ and a time-dependent term $I_R(x,y,t)$ with
\begin{eqnarray}
\label{micro_rotor} 
I_L(x,y) &\equiv& (A_1+A_2)^2 |\phi(x,y)|^2\, ,
\nonumber \\  \\   \nonumber
I_R(x,y,t) &\equiv& 2 A_1A_2 |\phi(x,y)|^2 \cos\left(2S(x,y) - \Omega t\right).
\end{eqnarray}
Both terms give rise to light shift potentials according to $V_{L,R} = - \Re(\alpha) I_{L,R}$ with atomic polarizability $\alpha$ \cite{Gri:00}. The amplitudes $A_1$ and $A_2$ can be chosen to control the ratio $I_R/I_L$. We restrict ourselves to $\sin(\theta) = 1$ in the following. In this case, the stationary term provides a square lattice $I_L(x,y) = (A_1+A_2)^2 [\sin^2(kx)+\sin^2(ky)]$, while the time-dependent term $I_R(x,y,t)$ acts as a collection of microscopic rotors, which apply angular momentum with alternating sign to the $\lambda/2\times \lambda/2 $-sized plaquettes of the square lattice $I_L(x,y)$. As illustrated in Fig.\ (2), in the vicinity of each node of $I_L(x,y)$ the term $I_R(x,y,t)$ provides a rotating quadrupole potential with alternating sense of rotation for adjacent lattice sites. In Fig.\ 2(b)-(j) the micro-rotor intensity $I_R(x,y,t)$ is shown for $\Omega t = n \pi/8$, with $n \in \{0,1,...,8\}$, thus illustrating a $\tilde \Omega \, t = \pi/2$ clockwise rotation of the quadrupole around the central node of $I_L(x,y)$ in Fig.\ 2(a). Hence, the angular frequency of the micro-rotors is $\tilde \Omega = \Omega /2$. 

\begin{figure}
\includegraphics[scale=0.28]{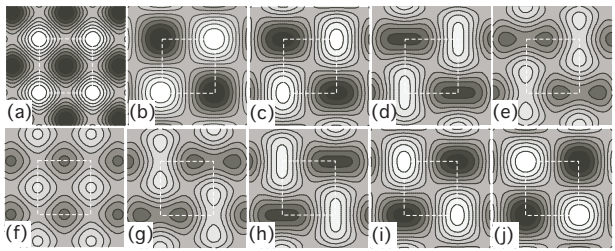}
\caption{\label{fig.2} In (a) the square lattice intensity $I_L(x,y)$ is plotted. White (black) regions indicate antinodes (nodes), which for negative detuning correspond to potential minima. In (b)-(j) the micro-rotor intensity $I_R(x,y,t)$ is shown (same area as in (a)) for $\Omega t = n \pi/8$ with $n \in \{0,1,...,8\}$. The grey scale indicates values of $I_R(x,y,t) / 4A_1A_2$ between $-1$ (black) to $+1$ (white). The dashed white squares indicate the position and size of the plaquettes.}
\end{figure}

Temporary application of the micro-rotor potential $V_{R}(r,t)$ may be used to resonantly excite the $d_{x^2-y^2}$-state $\psi$ by adjusting $\hbar \Omega$ equal to the energy difference between $\psi$ and the ground state of the lattice \cite{Yuk:02}. Let us briefly estimate the resonance condition. The angular momentum applied to each plaquette is approximately given by $m \, \tilde \Omega \, r^2$, where $r \equiv \lambda/4$ is the distance from the centre to the edge of the plaquette. Excitation of vortices requires an angular momentum of $\hbar$ per plaquette, i.e., $m \, \tilde \Omega \, r^2 \approx \hbar$ and thus $\hbar  \, \Omega \approx (8/ \pi^2) \, 2E_{\rm rec}$. For rubidium atoms and a convenient lattice wavelength ($\lambda$ = 1030 nm) $\Omega/2\pi = 3.5$ kHz. Since $\hbar \Omega$ turns out to be comparable to the collisional energy $g \bar \rho$, we cannot directly apply the theory of Ref. \cite{Yuk:02} to estimate the excitation efficiency. A detailed calculation is involved and beyond the scope of this article. Nevertheless, the appropriate transfer of angular momentum by $V_{R}(r,t)$ together with the metastable character of $\psi$ yield strong evidence that $\psi$ can in fact be efficiently excited by means of $V_{R}(r,t)$. 

Experimentally, the generation of the bichromatic light field of Eq.\ (\ref{micro_rotor}) is straightforward using the optical set-up illustrated in Fig.\ 3(a), thus extending a method proven practicable in previous experiments \cite{Experiment}. The two components $E_1( \mathbf r, t)$ and $E_2( \mathbf r, t)$  are produced in two nested Michelson-interferometers, each with its two branches folded under 90$^\circ$ angle. Two laser beams with adjustable frequency difference and linear polarization perpendicular to the drawing plane in Fig.\ 3(a) are used to couple both interferometers. Each interferometer comprises a piezo-electrically driven mirror, which permits to adjust the appropriate relative phase of the fields in each branch. In order to produce the required waves $\phi(x,y) = e^{i \pi/4}\sin(kx) + e^{-i \pi/4}\sin(ky)$ and $\phi^{*}(x,y) = e^{-i \pi/4}\sin(kx) + e^{i \pi/4}\sin(ky)$, the optical path length differences are set to $+\lambda/4$ and $-\lambda/4$.

\begin{figure}
\includegraphics[scale=0.24]{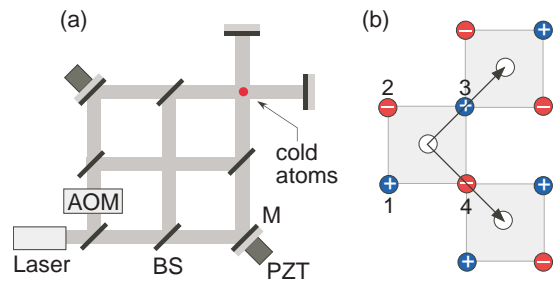}
\caption{\label{fig.3} (a) Optical set-up consisting of two nested Michelson-interferometers. PZT = piezo-electric transducer, M = mirror, BS = beam splitter, AOM = acousto-optic frequency shifter. (b) The lattice is decomposed into plaquettes (grey rectangles) with potential minimas in the corners (filled disks) and maximas in the centers (white disks) translated along the basis vectors indicated by the black arrows.}
\end{figure}

Finally, we wish to point out that the micro-rotor term $I_R(x,y,t)$ gives rise to driven ring-like tunneling in the Bose-Hubbard model. Expanding the boson field operator in terms of the Wannier-function $W(r)$ of the first band yields
\begin{eqnarray}
\label{BH_interaction1} 
H_R \equiv & \int & dr \, \Psi^{+}(r,t) V_{R}(r,t) \Psi(r,t)  = 
\nonumber \\  \\   \nonumber
 \sum_{n, m}\,\, b_{n}^{+} b_{m}  \, & \int & dr \, W^{*}(r-r_n) V_{R}(r,t) 
W(r-r_m) \,,
\end{eqnarray} 
where $V_{R}(r,t) \equiv V(r) e^{-i \Omega t} \, + \,  \textnormal{c.c.}$ is the micro-rotor potential ($dr \equiv dxdy$) and $n,m$ are summed over all lattice sites. This yields $H_R\, =\, \sum_{n,m}\,\, b_{n}^{+} b_{m} \, K_{nm} \, e^{-i \Omega t} \,\, + \,\,  \textnormal{H.c.} $ with $K_{nm} \equiv \int dr W^{*}(r-r_n) V(r) W(r-r_m)$. Keeping only diagonal and nearest neighbour terms and assuming that $W^{*}(r-r_n)W(r-r_m) \neq 0$ mainly around $(r_n+r_m)/2$, one may write $K_{nm} = \kappa_{nm} \, V((r_n+r_m)/2)$ with $\kappa_{nm} \equiv \int dr W^{*}(r-r_n) W(r-r_m)$ satisfying $\kappa_{nm} = \kappa_{mn} = K \in \mathbb{R}$ for $n \neq m$ and $\kappa_{nn} =  1$. The lattice can be composed by translations of $\lambda/2 \times \lambda/2$-sized square plaquettes (grey rectangles in Fig.\ 3(b)) along the Bravais-lattice $\mathcal{R} \equiv \{ R_{\nu\mu} = \nu R_{+} + \mu R_{-} | \nu, \mu \in \mathbb{Z}\}$ with basis vectors $R_{\pm} \equiv (\lambda/2) (\hat x \pm \hat y)$, respectively (black arrows in Fig.\ 3(b)). Assume that the four potential minima at each corner of some plaquette (filled disks in Fig.\ 3(b)) are indicated by $n = 1,2,3,4$ starting from the lower left corner in clockwise order. Using $V(r) = (V_{1}/2) \, \phi^2(x,y)$ with a real constant $V_{1}$, the coupling constants are evaluated as $K_{12} = K_{34} = -K_{23} = -K_{41} = i V_{1} K/2$ and $K_{11} = K_{33} = -K_{22} = -K_{44} = V_{1}$. Summing over all plaquettes finally yields  
\begin{eqnarray}
\label{BH_interaction2} 
H_R  \, &=&  \, V_{1} K \sin(\Omega t) \sum_{\mathcal{R}} \,(b_{4}^{+} - b_{2}^{+})(b_{3} - 
b_{1})  \, \, + \, \,{\rm  H.c.} 
\nonumber \\  &+& \, 2 V_{1}  \cos(\Omega t)\, (\hat N_{+}-\hat N_{-}) \,\, ,
\end{eqnarray}
where the operators $\hat N_{+} \equiv (1/2)\sum_{\mathcal{R}} \, b_{1}^{+}b_{1} +b_{3}^{+}b_{3}$ and $\hat N_{-} \equiv (1/2)\sum_{\mathcal{R}} \, b_{2}^{+}b_{2} +b_{4}^{+}b_{4}$ denote the total particle numbers in the two sublattices indicated by $(+)$ and $(-)$ signs in Fig.\ 3(b). 

The Hamiltonian $H_R$ has to be added to the conventional Bose-Hubbard Hamiltonian $H_{\rm BH}$ describing the optical lattice in absence of $V_{R}(r,t)$ according to Ref.\ \cite{Jak:98}. The first term comprises next neighbour hopping terms oscillating in phase for opposite edges of a plaquette and with opposite phase for adjacent edges. The second term provides potentials oscillating in phase for lattice sites on the same diagonal and with opposite phase for adjacent sites. Hopping and potential terms exhibit $90^\circ$ phase lag, thus giving rise to a rotational direction of $H_R$, which suggests the interpretation of a driven ring-like tunneling process. Note that $H_R$ is quadratic with respect to the particle operators, in contrast to the quartic ring exchange interactions considered in Refs.\ \cite{Buc:05, ring_exchange}. The ground state of $H_{\rm BH}$ exhibits a well known phase diagram possessing a Mott insulator and a superfluid phase \cite{Fis:89}. The implementation of $V_{R}(r,t)$ allows to experimentally study the phase diagram of $H_{\rm BH} + H_R$. Whether it provides additional quantum phases (e.g. a normal Bose liquid, which is compressible but not superfluid) is an exciting question to be addressed. Via temporary application of $H_R$ one could explore the possible existence of an excited state of $H_{\rm BH}$ corresponding to the mean field solution given by Eq.\ (\ref{solution}).
 
In summary, we have discussed the existence of a $d_{x^2-y^2}$-density wave in a two-dimensional bosonic optical lattice with square lattice geometry. We have proposed a straightforward experimental scenario to excite this state, which in the Bose-Hubbard picture amounts to the insertion of anisotropic time-dependent hopping and potential terms. This opens up a new axis of phase space, possibly giving rise to new exciting quantum phases. Our considerations are not constrained to square lattice potentials. For example, the superposition of six light beams travelling in the $xy$-plane (one every $60^{\circ}$ with alternating frequencies) yields a similar scenario of alternating rotational flux, however with hexagonal lattice symmetry. In this article, we have restricted ourselves to bosonic atoms, however, similar scenarios involving fermions, mixed systems or molecules appear possible. 

\begin{acknowledgments}
AH acknowledges support by DFG (He2334/6-2). We thank A. O. Caldeira, R. L. Doretto, L.- K. Lim, and C. Zimmermann for their helpful remarks. 
\end{acknowledgments}


\begin{thebibliography}{21}

\bibitem{Aff:88}
I. Affleck and J. B. Marston, Phys. Rev. B {\bf 37}, 3774 (1988); J. B. Marston and I. Affleck, Phys. Rev. B {\bf 39}, 11538 (1989).
\bibitem{Zha:90}
F. C. Zhang, Phys. Rev. Lett. {\bf 64}, 974 (1990).
\bibitem{Hsu:91}
T. Hsu {\em et~al.}, Phys. Rev. B {\bf 43}, 2866 (1991).
\bibitem{Nay:00}
C. Nayak, Phys. Rev. B {\bf 62}, 4880 (2000).
\bibitem{Cha:01}
S. Chakravarty {\em et~al.}, Phys. Rev. B {\bf 63}, 094503 (2001).
\bibitem{QuantumGases}
K. Burnett {\em et~al.}, Nature (London) {\bf 416}, 225 (2002); J. Anglin and W. Ketterle, Nature (London) {\bf 416}, 211 (2002).
\bibitem{Lew:06}
M. Lewenstein {\em et~al.}, cond-mat/0606771 (2006).
\bibitem{Fis:89}
M. Fisher {\em et~al.}, Phys. Rev. B {\bf 40}, 546 (1989). 
\bibitem{Jak:98}
D. Jaksch {\em et~al.}, Phys. Rev. Lett. {\bf 81}, 3108 (1998).
\bibitem{Gre:02}
M. Greiner {\em et~al.}, Nature (London) {\bf 419}, 51 (2002).
\bibitem{Molecules}
T. St\"oferle {\em et~al.}, Phys. Rev. Lett. {\bf 96}, 030401 (2006), C. Ospelkaus {\em et~al.}, Phys. Rev. Lett. {\bf 97}, 120402 (2006).
\bibitem{Koe:06}
A. Koetsier {\em et~al.}, Phys. Rev. A {\bf 74}, 033621 (2006).
\bibitem{QuantumHall}
D. Jaksch and P. Zoller, New J. Phys. {\bf 5}, 56 (2003); A. S\o rensen {\em et~al.}, Phys. Rev. Lett. {\bf 94}, 086803 (2005); R. N. Palmer and D. Jaksch, Phys. Rev. Lett. {\bf 96}, 180407 (2006).
\bibitem{San:04}
L. Santos {\em et~al.}, Phys. Rev. Lett. {\bf 93}, 030601 (2004).
\bibitem{Buc:05}
H. B\"uchler {\em et~al.}, Phys. Rev. Lett. {\bf 95}, 040402 (2005).
\bibitem{Rei:05}
J. W. Reijnders and R. A. Duine, Phys. Rev. A. {\bf71}, 063607 (2005).
\bibitem{Tun:06}
S. Tung {\em et~al.}, Phys. Rev. Lett. {\bf97}, 240402 (2006).
\bibitem{Hem:91}
A. Hemmerich {\em et~al.}, Phys. Rev. A. {\bf 44}, 1910 (1991).
\bibitem{VortexExcitation}
B. Jackson {\em et~al.}, Phys. Rev. Lett. {\bf 80}, 3903 (1998); K.-P. Marzlin {\em et~al.}, Phys. Rev. Lett. {\bf 79}, 4728 (1997); K.-P. Marzlin and W. Zhang, Phys. Rev. A. {\bf57}, 3801 (1998).
\bibitem{Gri:00}
R. Grimm {\em et~al.}, Adv. At., Mol., Opt. Phys. {\bf 42}, 95 (2000).
\bibitem{Yuk:02}
V. I. Yukalov {\em et~al.}, Phys. Rev. A. {\bf66}, 043602 (2002).
\bibitem{Experiment}
A. Hemmerich {\em et~al.}, Europhys. Lett. {\bf 18}, 391 (1992); A. Hemmerich and T. W. H\"{a}nsch, Phys. Rev. Lett. {\bf 68}, 1492 (1992); A. Hemmerich {\em et~al.}, Europhys. Lett. {\bf 21}, 445 (1993). 
\bibitem{ring_exchange}
S. Brehmer {\em et~al.}, Phys. Rev. B. {\bf 60}, 329 (1999); O. I. Motrunich and T. Senthil, Phys. Rev. Lett. {\bf 89}, 277004 (2002); U. Schollw\"ock {\em et~al.}, Phys. Rev. Lett. {\bf 90}, 186401(2003); V. Rousseau {\em et~al.}, Phys. Rev. B {\bf 72}, 054524 (2005); A. M. Toader {\em et~al.}, Phys. Rev. Lett. {\bf94}, 197202 (2005).

\end{thebibliography}
\end{document}